# Zero-order-free complex beam shaping

Yansheng Liang[1], Xue Yun[1], Minru He[1], Zhaojun Wang[1], Ming Lei[1,*]

[1] *MOE Key Laboratory for Nonequilibrium Synthesis and Modulation of Condensed Matter, Shaanxi Province Key Laboratory of Quantum Information and Quantum Optoelectronic Devices, School of Physics, Xi'an Jiaotong University, 710049, China*
*ming.lei@mail.xjtu.edu.cn

**The unwanted zero-order diffraction is still an issue in beam shaping using pixelated spatial light modulators. In this paper, we report a new approach for zero-order free beam shaping by designing an asymmetric triangle reflector and introducing a digital blazed grating and a digital lens to the phase hologram addressed onto the spatial light modulator. By adding the digital lens phase to the previously reported complex-amplitude coding algorithms, we realized the generation of complex beams without the burden of zero-order diffraction. We comparatively investigated the produced complex light fields using the modified complex-amplitude coding algorithms to validate the proposed method.**

http://dx.doi.org/10.1364/xxx.99.099999

Beam shaping aims to create optical beams with desired polarization, intensity, and phase distribution, which is a long-term issue of high interest in science and technology [1-3]. In the past two decades, the development of beam shaping has been significantly advanced, from generating simple symmetric intensity patterns [4, 5] to shaping light fields with arbitrary three-dimensional trajectories [6-9] and desired polarization distribution [10].

Spatial light modulators (SLMs) are pixelated devices allowing the efficient shaping of light into arbitrary patterns by optimized computer-generated holograms (CGHs) [11]. However, even if the CGH is ideally designed, the un-diffracted component exists because of the pixelated structure of the SLM, which is a typical problem in beam shaping using SLMs [12]. This un-diffracted component is the so-called zero-order. Although it takes only a small fraction of the illumination power, it is non-ignorable in many cases, such as optical trapping [13, 14], microscopy [15], etc. Besides the pixelated structure, the design of the CGHs also will lead to the zero-order. For example, the power of the zero-order can be much higher than the diffracted component in complex beam shaping [16, 17].

For the zero-order arising from the pixelated structure, efforts have been made to suppress it. The direct method deviates the modulated component away from the zero-order with a blazed grating and places a beam block at the focal plane of a lens to remove the zero-order [18]. Alternatively, illuminating the SLM with a diverging or converging beam can axially shift the zero-order away from the modulated component [19]. However, there still leaves a small part of the zero-order mixed with the modulated component. The interference method is also employed to suppress the zero-order. By adding a destructively interfering spot coinciding with the zero-order, the power of the zero-order can be significantly reduced [20, 21]. However, the suppression efficiency depends highly on the addressed phase hologram.

The abovementioned methods have been working well in many applications of beam shaping. However, considering their drawbacks, they are not suitable for complex beam shaping techniques based on complex-amplitude coding algorithms, as the power of the zero-order can be very high and highly depends on the design of the CGH. In this paper, we proposed a new approach for beam shaping without the burden of the zero-order based on a digital lens and an asymmetric triangle reflector. We employed this approach to complex beam shaping by introducing the digital lens phase to the complex-amplitude coding algorithms. By doing so, we realized zero-order free complex beam shaping.

For the beam shaping configuration using an SLM, the ideal illumination angle of the input beam on the SLM is zero degrees (Fig. 1(a))). In this way, the beam shaping resolution is high, but the power loss is >75% because of a cube non-polarizing beamsplitter [22]. Therefore, small-angle illumination is generally adopted (Fig. 1(b)), requiring the incidence angle as small as possible for high-resolution modulation, e. g., <10 degrees. In such a system, a lateral shift to the modulated component (the +1$^{st}$ order) is first applied by adding a blazed grating to the hologram (Fig. 1(c)). Then, a spatial filter is placed at the focal plane of lens1 to block the zero-order. However, it is not suitable for complex beam shaping because of the high-power zero-order.

Vertical illumination but small-angle output might be a solution (Fig. 1(d)). In this configuration, the zero-order will be reflected along the inverse direction of the input beam, and the modulated component will be separated from the zero-order by adding a blazed grating to the hologram. For the light wavelength of $\lambda$, the output angle $\theta$ satisfies the relation $\sin\theta=\lambda/\Lambda$ with $\Lambda$ denotes the phase period of the blazed grating. Setting $\lambda$=1.064 μm, we get $\Lambda$ of ~10.2 μm for $\theta$=6 degrees and ~20.3 μm for $\theta$=3 degrees. Unfortunately, the pixel pitch of the commercial SLM is ~8 um for 1080p resolution and ~3.6 μm for 2k resolution. Therefore, this method is not practical for using a 1080p SLM. For a 2K SLM, $\Lambda$ is ~6 pixels, which may satisfy the demand of the beam shaping resolution. However, the distance for separating the input beam with a diameter of 8.5 mm and the output beam, i.e., the vertical distance from the separation point O to the SLM, is ~162.4 mm. Considering the size of the optical elements used in the following optical path, e.g., the 4f system consisting of lenses 1 and 2 in Fig. 1(d), the separation distance would be much longer.

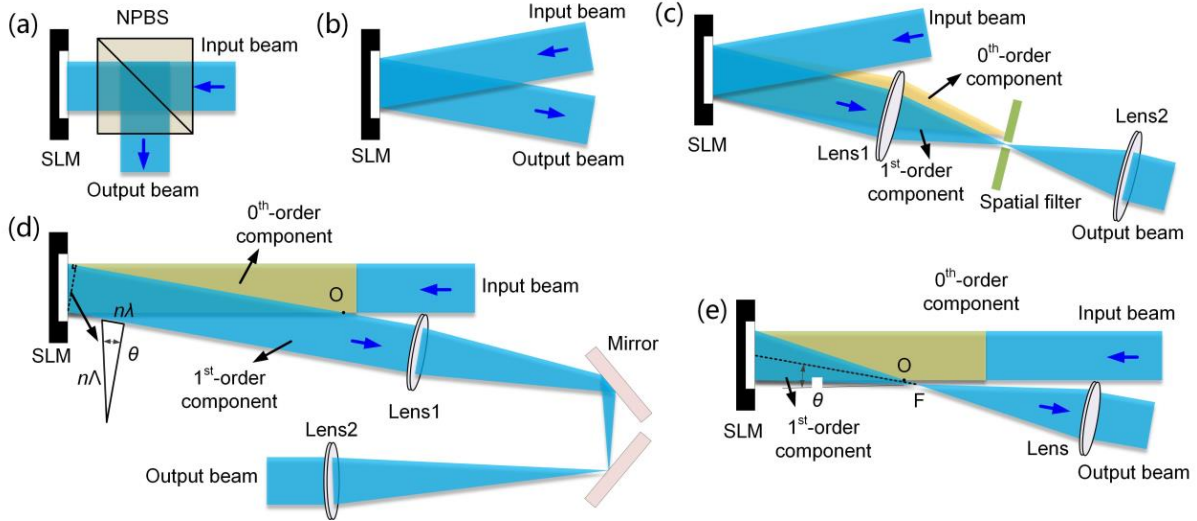

Figure 1 The working model of a reflective SLM and the principle of suppressing the zero-order diffraction in beam shaping. (a, b) Normal incidence (a) and small-angle incidence of the beam on the SLM. (c) Blocking the zero-order by a spatial filter in the system employing small-angle incidence of the beam on the SLM. (d) The zero-order-free system based on normal incidence but small-angle output by addressing a blazed grating onto the SLM. (d) The zero-order-free system based on normal incidence but small-angle output by simultaneously addressing a blazed grating and a digital lens onto the SLM.

We further advanced the configuration shown in Fig. 1(d) by introducing a digital lens to the CGH in addition to a blazed grating (Fig. 1(e)). The minimal distance separating the input and output beams will be achieved when the foci of the digital lens and the separation point O coincide. Similarly, for $\lambda$=1.064 μm and $\theta$=1.5 degrees, $\Lambda$ is ~40.6 μm, about 5 pixels for a 1080p SLM, which is practical to realize. In this case, the minimum separation distance is ~81.2 mm for the input beam diameter of 8.5 mm. The advanced configuration shown in Fig. 1(e) permits vertical illumination, a smaller output angle, and a shorter separation distance.

In the following, we will validate the efficiency of our method in beam shaping with simulations (Fig. 2). As an example, our goal was to create a perfect optical vortex (POV), whose ring size is independent of the topological charge (TC) [23-25]. The input beam is a Bessel beam with TC=10 and complex amplitude $E_{in}=A_{Bessel}\cdot\exp(i\phi_{Bessel})$, where $A_{Bessel}$ and $\phi_{Bessel}$ denoting the phase (Fig. 2(a)) and amplitude (Fig. 2(b)), respectively. By Fourier transforming the Bessel beam, we obtained a POV in the focal plane (Fig. 2(c)). Now let us find out whether the vortex generated at the focal plane of the digital Fresnel lens is perfect or not. According to the Fresnel diffraction integral, the optical field propagates after a distance of $z$ satisfies the relation [26]

$$E(x',y')=\frac{1}{i\lambda z}\iint E_{in}(x,y)\cdot e^{ik[(x'-x)^2+(y'-y)^2]/2z}\mathrm{d}x\mathrm{d}y. \qquad (1)$$

Here, $(x,y,z)$ and $(x',y')$ denote the Cartesian coordinates. We used the Fresnel lens as the digital lens, which is ritten as

$$\phi_{lens}(x,y)=\frac{2\pi(x^2+y^2)}{\lambda f}, \qquad (2)$$

where $f$ denotes the focal length of the lens. The input beam after adding a digital lens to its phase, is turned into $E_{in}=A_{Bessel}\cdot\exp[i(\phi_{Bessel}+\phi_{lens})]$, of which the phase distribution is shown in Fig. 2(d). Based on Eq. (1), we obtained the phase and intensity of the light field propagating to the focal plane of the digital lens ($z$=150 mm) shown in Figs. 2(e) and 2(f), respectively. The azimuthal phase change period is 10, indicating that the TC is 10, equivalent to the TC of $E_{in}$. To separate the modulated beam and the zero-order, we further introduce a blazed grating, expressed by

$$\phi_g(x,y) = \mathrm{mod}(\frac{2\pi x}{\Lambda}, 2\pi) \,. \tag{3}$$

The phase of $E_{in}$ after adding a blazed grating with $\Lambda$=16 pixels is shown in Fig. 2(g). Similarly, we obtained the shifted vortex intensity (Fig. 2(h)) and its spiral phase distribution (Fig. 2(i)) at $z$=150 mm. The spiral phase distribution has an azimuthal period of 10, equivalent to the TC of the input Bessel beam. The co-axial interference pattern by two shifted vortices with the same ring size but various TC (for one TC =10 and the other TC=0) shows 10 fringes, confirms that the TC of the vortex is 10 (Figs. 2(j)). Further analysis in Fig. 2(k) demonstrates that the measured TC is always equal to the TC of the input Bessel beam. Moreover, the ring size keeps unchanged with increasing TC, indicating the generated vortex is "perfect" (Fig. 2 (l)). Therefore, the POV and other complex beams can be generated at the focal plane of the digital lens.

For the generation of a complex beam, simultaneous shaping of the amplitude and phase of the input beam is generally needed. Approximate complex-amplitude modulation can be realized with a single phase-only SLM by encoding the target field's amplitude and phase information in a phase hologram [27-29]. Here, we advanced the commonly employed amplitude-coding algorithm by adding a digital lens to the modulated phase. By doing so, we can realize zero-order-free complex beam shaping.

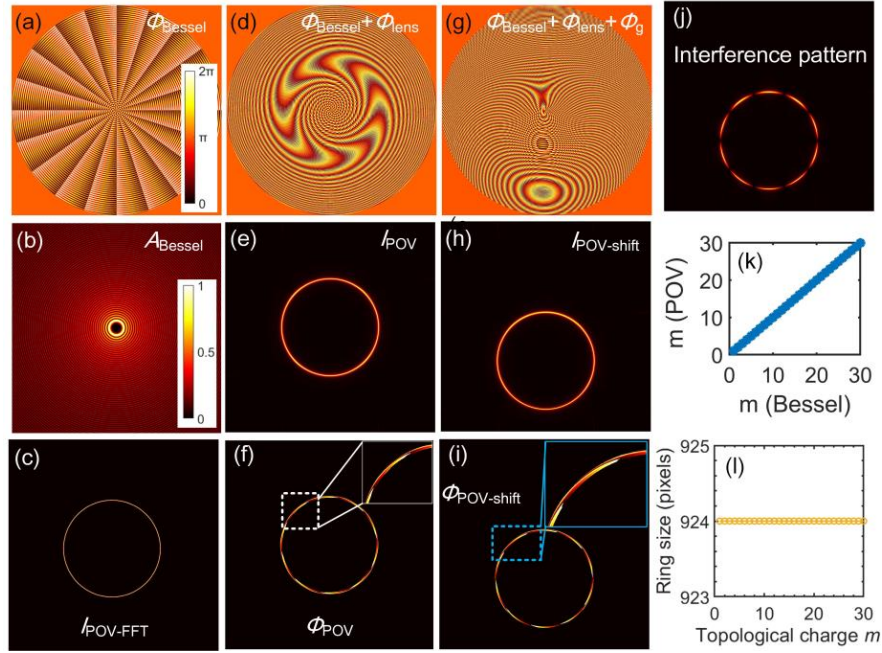

Figure 2 Simulations for generating a POV with a digital lens. (a, b) The phase and amplitude of a Bessel beam with TC=10. (c) The intensity of a POV generated by the Fourier transform of a Bessel beam shown in parts (a) and (b). (d) The phase calculated from the combination of a lens and a Bessel beam. (e, f) The intensity and phase of the vortex at the focal plane of the digital lens. (g) The phase calculated from the combination of a lens, a blazed grating, and a Bessel phase. (h, i) The intensity and phase of the shifted vortex at the focal plane of the digital lens. (j) The interference pattern. (k, l) The TC of the POV generated at the focal plane of a digital lens and the ring size of the vortex against its TC. Insets in parts (f) and (i) present the enlarged field of view of the phase distribution.

By using a phase hologram to shape the input beam $E_{in}(x,y)=A_{in}(x,y)\cdot\exp(i\phi_{in}(x,y))$ to the target field $E_{tar}(x,y)=A_{tar}(x,y)\cdot\exp(i\phi_{tar}(x,y))$, these two fields satisfy the relation [30]

$$E_{in}(x,y)e^{i\mathbf{k}_{in}\cdot\mathbf{r}} \cdot e^{iH(x,y)} = E_{tar}(x,y)e^{i\mathbf{k}_{tar}\cdot\mathbf{r}} \,. \tag{4}$$

Here, $\mathbf{r}=(\mathbf{x,y,z})$, $H(x,y)$ is the phase imprinted to the input beam, $\mathbf{k}_{in}$ and $\mathbf{k}_{des}$ are the wave vectors of the input and the target fields, respectively. Let $A_{rel}=A_{tar}(x,y)/A_{in}(x,y)$ and $\phi_{relg}(x,y)=\phi_{tar}(x,y)-\phi_{in}(x,y)+(\mathbf{k}_{tar}-\mathbf{k}_{in})\cdot\mathbf{r}$, we get

$$e^{iH(x,y)} = A_{rel}(x,y)e^{i\phi_{relg}(x,y)} \,. \tag{5}$$

Equation (5) can be modified to

$$H(x,y) = F(A(x,y))\phi_{relg}(x,y) \,, \tag{6}$$

where $F(\cdot)$ denotes the function to encode the complex amplitude to a phase hologram. Setting $(\mathbf{k}_{tar}-\mathbf{k}_{in})\cdot\mathbf{r}$ as the phase of a blazed grating, $\phi_g(x,y)$, the relative phase is written as

$$\phi_{\text{relg}}(x, y) = \phi_{\text{tar}}(x, y) + \phi_{\text{in}}(x, y) + \phi_{\text{g}}(x, y) \, . \tag{7}$$

Many complex-amplitude coding algorithms have been reported for finding F(·) [27-29]. To modify the amplitude-coding algorithm, we further introduce a digital lens to $\phi_{\text{relg}}$, which is then modified to

$$\phi_{\text{relg}}(x, y) = \phi_{\text{tar}}(x, y) - \phi_{\text{in}}(x, y) + \phi_{\text{g}}(x, y) + \phi_{\text{lens}}(x, y) \, . \tag{8}$$

Based on Eqs. (5) and (8), the desired complex field can be generated in the focal region of the digital lens.

The experimental setup for realizing the zero-order-free complex beam shaping is shown in Fig. 3. The linearly polarized beam (wavelength 532 nm, maximal power 2 W) first passed through the beam isolator consisting of a cube polarization beamsplitter, a Faraday rotator, and a half-wave plate (HWP). The HWP was used to transfer the beam polarization to the $p$-direction. Then the beam was expanded by Lenses 1 and 2 before being modulated by a pure-phase SLM (Pluto-NIR-II, HOLOEYE Photonics AG Inc., Germany), which had a resolution of 1920×1080 pixels and pixel pitch of 8 μm. A specially designed asymmetric triangle reflector was employed to reflect the input collimated beam onto the SLM and then guided the retro-reflected modulated beam to Lense 3. The camera and the foci of the digital lens were located at the 2$f$ positions of Lens 3.

If the basic angle of the triangle reflector at the input surface is set to 45°, the input beam will illuminate the SLM panel vertically. By imprinting a blazed grating onto the SLM, the modulated beam will be output with a small angle $\theta$. In contrast, the zero-order will be reflected backward and then pass through the beam isolator before being blocked. Another basic angle $\beta$ of the reflector satisfies the following relation: $\beta=\theta/2$. Consequently, the beam (Arrow 2) reflected by the reflector has the same direction as the input beam (Arrow 1). In our experiment, the output angle $\theta$ is set to ~1.5 degrees, corresponding to $\Lambda$=5 pixels, while $f$ is set as 180 mm.

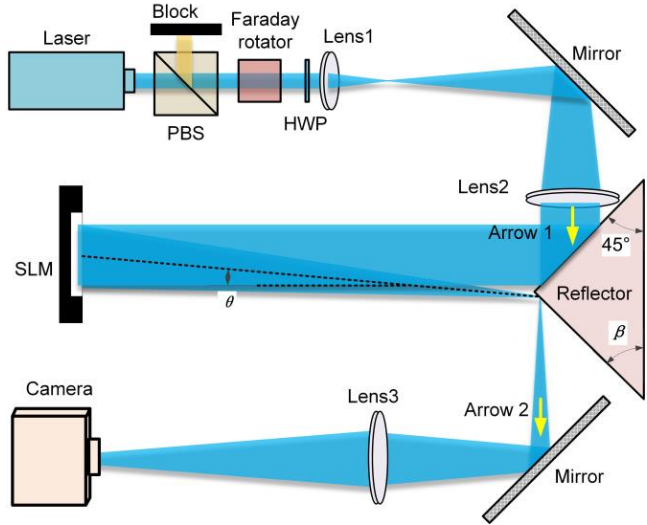

Figure 3 The experimental setup for realizing the zero-order-free complex beam shaping. PBS: polarizing beam-splitter; HWP: half-wave plate.

To verify the efficiency of our method in the experiment, we modified the complex-amplitude coding algorithms reported by Arrizon [27] by introducing the digital lens to the phase hologram. Then the amplitude-coded phase holograms calculated with the modified algorithms were addressed onto the SLM. The experimental results are shown in Fig. 4. Figures 4(a1)-4(b1) (Column 1 of Fig. 4) show the experimentally generated circular (POV), line, and square patterns with TC=10. The input complex amplitude of the polymorphic beams was calculated using Rodrigo et al.'s method [7, 31]. All the generated patterns show high quality without the burden of the zero-order. Columns 2-4 present the interference pattern for determining the TC of the created vortex beams. The interference patterns were obtained by interfering the target beam with another polymorphic beam with the same trajectory but TC=0. From the figures, we can find that the interference patterns shown in Columns 2-4 in Fig. 4 have 1, 10, and 20 fringes, equivalent to the designed TC. Therefore, we have produced the desired complex fields with designed intensity patterns and phase distribution by using our configuration based on an asymmetric triangle reflector and the modified complex-amplitude-coding algorithm based on a digital lens.

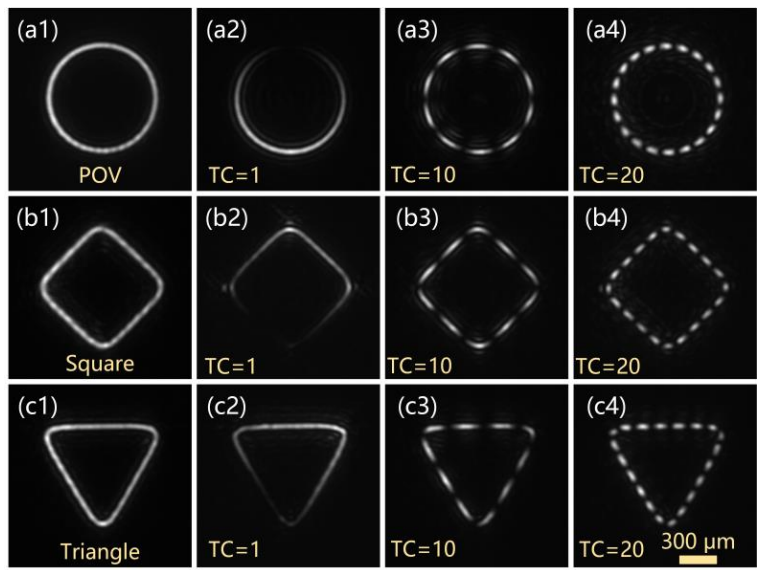

Figure 4 Experimental results of various complex light fields produced using the modified complex-amplitude coding algorithms and the interference patterns for determining the TC. (a1, b1, c1) The intensity patterns of the created polymorphic beams with (a1) circular, (b1) square, and (c1) triangle shape and TC=10 based on the modified complex-amplitude coding algorithm. (a2-a4), (b2-b4), and (c2-c4) The interference patterns for determining the TC of (a2-a4) circular-, (b2-b4) square-, and (c2-c4) triangle-shape polymorphic beams. Columns 2-4 refer the TC of 1, 10, and 20.

In summary, we have reported a new approach for zero-order free beam shaping. We proposed a method for vertical illumination but small-angle output of the beam by designing an asymmetric triangle reflector and introducing a digital blazed grating and a digital lens to the phase hologram. We advanced the commonly employed amplitude-coding algorithms to realize zero-order-free complex beam shaping. We demonstrated the efficiency of the reported method and configuration by simulations and experimental measurements. The results show that the target complex light fields can be generated at the digital lens's focal plane without the zero-order burden. We believe our method and configuration will have promising applications employing beam shaping, especially those using high-power lasers, such as laser inscribing.

**Funding**. National Natural Science Foundation of China (NSFC) (61905189, 62135003, 62005208); The Innovation Capability Support Program of Shaanxi (Program No. 2021TD-57); China Postdoctoral Science Foundation (2019M663656, 2020M673365).

**Disclosures.** The authors declare no conflicts of interest.

**Data availability.** Data underlying the results presented in this paper are not publicly available at this time but may be obtained from the authors upon reasonable request.